\documentclass[twocolumn,preprintnumbers,amsmath,amssymb,floatfix]{revtex4}

\usepackage{graphicx}
\usepackage{dcolumn}
\usepackage{bm}

\begin{document}

\title{System size dependence of cluster properties from two-particle angular correlations in Cu+Cu and Au+Au collisions at \boldmath $\sqrt{s_{_{NN}}}$ = 200~GeV}
\author{
%
B.Alver$^4$,
B.B.Back$^1$,
M.D.Baker$^2$,
M.Ballintijn$^4$,
D.S.Barton$^2$,
R.R.Betts$^6$,
R.Bindel$^7$,
W.Busza$^4$,
Z.Chai$^2$,
V.Chetluru$^6$,
E.Garc\'{\i}a$^6$,
T.Gburek$^3$,
K.Gulbrandsen$^4$,
J.Hamblen$^8$,
I.Harnarine$^6$,
C.Henderson$^4$,
D.J.Hofman$^6$,
R.S.Hollis$^6$,
R.Ho\l y\'{n}ski$^3$,
B.Holzman$^2$,
A.Iordanova$^6$,
J.L.Kane$^4$,
P.Kulinich$^4$,
C.M.Kuo$^5$,
W.Li$^4$,
W.T.Lin$^5$,
C.Loizides$^4$,
S.Manly$^8$,
A.C.Mignerey$^7$,
R.Nouicer$^2$,
A.Olszewski$^3$,
R.Pak$^2$,
C.Reed$^4$,
E.Richardson$^7$,
C.Roland$^4$,
G.Roland$^4$,
J.Sagerer$^6$,
I.Sedykh$^2$,
C.E.Smith$^6$,
M.A.Stankiewicz$^2$,
P.Steinberg$^2$,
G.S.F.Stephans$^4$,
A.Sukhanov$^2$,
A.Szostak$^2$,
M.B.Tonjes$^7$,
A.Trzupek$^3$,
G.J.van~Nieuwenhuizen$^4$,
S.S.Vaurynovich$^4$,
R.Verdier$^4$,
G.I.Veres$^4$,
P.Walters$^8$,
E.Wenger$^4$,
D.Willhelm$^7$,
F.L.H.Wolfs$^8$,
B.Wosiek$^3$,
K.Wo\'{z}niak$^3$,
S.Wyngaardt$^2$,
B.Wys\l ouch$^4$\\
\vspace{3mm}
\small
%
%
%
%
$^1$~Argonne National Laboratory, Argonne, IL 60439-4843, USA\\
$^2$~Brookhaven National Laboratory, Upton, NY 11973-5000, USA\\
$^3$~Institute of Nuclear Physics PAN, Krak\'{o}w, Poland\\
$^4$~Massachusetts Institute of Technology, Cambridge, MA 02139-4307, USA\\
$^5$~National Central University, Chung-Li, Taiwan\\
$^6$~University of Illinois at Chicago, Chicago, IL 60607-7059, USA\\
$^7$~University of Maryland, College Park, MD 20742, USA\\
$^8$~University of Rochester, Rochester, NY 14627, USA\\
}

\begin{abstract}
\noindent

We present results on two-particle angular correlations in Cu+Cu and Au+Au
collisions at a center of mass energy per nucleon pair of 200~GeV over a broad range 
of pseudorapidity ($\eta$) and azimuthal angle ($\phi$) as a function of collision centrality. 
The PHOBOS detector at RHIC has a uniquely-large angular coverage for inclusive charged 
particles, which allows for the study of correlations on both long- and short-range 
scales. A complex two-dimensional correlation structure in $\Delta \eta$ and $\Delta \phi$ 
emerges, which is interpreted in the context of a cluster model. The effective cluster 
size and decay width are extracted from the two-particle pseudorapidity correlation 
functions. The effective cluster size found in semi-central Cu+Cu and Au+Au collisions is 
comparable to that found in proton-proton collisions but a non-trivial decrease of the size with 
increasing centrality is observed. Moreover, a comparison between results from Cu+Cu and Au+Au 
collisions shows an interesting scaling of the effective cluster size with the measured fraction 
of total cross section (which is related to the ratio of the impact parameter to 
the nuclear radius, $b/2R$), suggesting a geometric origin. Further analysis 
for pairs from restricted azimuthal regions shows that the effective cluster size at $\Delta\phi \sim 180^{\circ}$ 
drops more rapidly toward central collisions than the size at $\Delta\phi \sim 0^{\circ}$. 
The effect of limited $\eta$ acceptance on the cluster parameters is also addressed, 
and a correction is applied to present cluster parameters for full $\eta$ coverage, 
leading to much larger effective cluster sizes and widths than previously noted 
in the literature. These results should provide insight into the hot and 
dense medium created in heavy ion collisions. 

\vspace{3mm}
\end{abstract}

\maketitle
\section{Introduction}
\label{intro}

Multiparticle correlation studies have played an important role in exploring 
the underlying mechanism of particle production in high energy hadronic collisions.
In p+p collisions, inclusive two-particle correlations have been found to 
have two components: ``intrinsic'' two-particle correlations as well as an effective ``long-range'' 
correlation due to event-by-event fluctuations of the overall particle multiplicity (see, for example, Ref.~\cite{UA5_3energy}).
By considering the two-particle rapidity density at fixed multiplicity, the intrinsic 
correlations between particles were isolated and found to be approximately Gaussian, 
with a range of $\sigma_{\eta} \sim 1$ unit in pseudorapidity ($\eta=-\ln(\tan(\theta/2))$). 
These correlations have been conventionally called ``short-range''. Their properties have been characterized
by the concept of ``cluster emission'' \cite{cluster_model}.

The simple idea that hadrons are produced in clusters, rather than individually, has had 
great success in describing many features of particle production \cite{UA5_3energy,cluster_model,ISR_twolowenergy,ISR_63GeV}. 
In this scenario, hadronization proceeds via ``clusters'', high mass states (e.g. resonances 
but not necessarily with all well-defined quantum numbers) which decay isotropically 
in their rest frame into the final-state hadrons. An independent cluster emission model (ICM) has been 
widely applied to the study of two-particle correlations
\cite{UA5_3energy,cluster_model,ISR_twolowenergy,ISR_63GeV,cluster_fit},
where clusters are formed before the final-state hadrons and are independently emitted according 
to a dynamically generated distribution in $\eta$ and $\phi$. The clusters subsequently decay 
isotropically in their own rest frame into the observed final-state hadrons.
The observed correlation strength and extent in phase space can be parameterized in 
terms of the cluster multiplicity, or ``size'' (the average number of particles in a cluster) 
and the decay ``width'' (which characterizes the separation of the particles in pseudorapidity). 
However, it should be noted that independent cluster emission is only a phenomenological approach which provides no 
insight as to the mechanisms by which clusters are formed. Further modeling is required to connect these 
studies to the underlying QCD dynamics. 

A measurement of cluster properties from two-particle correlations in p+p collisions 
for particles emitted into $|\eta|<$ 3 was performed previously at center of mass energies
($\sqrt{s}$) of 200~GeV and 410~GeV using the PHOBOS detector at the Relativistic Heavy 
Ion Collider (RHIC) \cite{phobos_pp}. The data suggested an effective cluster size ($K_{\rm eff}$, 
defined in Sect.~\ref{results}) of $2.44 \pm 0.08$ at
$\sqrt{s}$ = 200~GeV, increasing with collision energy and event multiplicity. 
The results are consistent with previous measurements from ISR and Sp\={p}S at various energies \cite{UA5_3energy,ISR_twolowenergy,ISR_63GeV}.

In heavy ion collisions at RHIC, it has been predicted that the formation 
of a strongly interacting quark gluon plasma could modify cluster properties 
relative to p+p collisions \cite{AAcluster_prediction}.  
In order to systematically explore the properties of the clusters from p+p to 
A+A collisions, this paper presents the results on two-particle angular correlations in Cu+Cu and 
Au+Au collisions at center of mass energy per nucleon pair ($\sqrt{s_{_{NN}}}$) of 200~GeV, 
over a very broad acceptance in $\Delta \eta$ ($ = \eta_{1} - \eta_{2}$) and $\Delta \phi$ 
($ = \phi_{1} - \phi_{2}$). The PHOBOS Octagon detector, covering pseudorapidity -3 $<\eta<$ 3 
over almost the full azimuth, is well suited to measure the correlations between particles 
emitted from clusters. The two-dimensional (2-D) correlation functions in Cu+Cu and Au+Au collisions, 
as well as the extracted effective cluster size and width from one-dimensional (1-D) $\Delta\eta$ 
correlation functions, are presented as a function of system size. By separating particle pairs into 
``near-side'' ($0^{\circ}<\Delta\phi<90^{\circ}$) and ``away-side'' ($90^{\circ}<\Delta\phi<180^{\circ}$), 
more detailed information on the properties of the clusters is obtained. 
Furthermore, extrapolating limited acceptance ($|\eta|<$ 3) to full phase space, cluster properties
unbiased by detector acceptance have been estimated. 
This comprehensive analysis of cluster properties in p+p and A+A collisions 
should provide useful information for understanding the hadronization stage, 
but may also give insight into physics relevant at much earlier times.


\section{DATA SETS}
\label{dataset}

The data presented here for Cu+Cu and Au+Au collisions at $\sqrt{s_{_{NN}}} =$ 200~GeV were collected 
during RHIC Run 4 (2004) and Run 5 (2005) using the large-acceptance PHOBOS Octagon silicon
array covering pseudorapidity $-3<\eta<3$ over almost the full azimuth. A full description of 
the PHOBOS detector can be found in Ref.~\cite{phobos_detector}. The primary event trigger
used the time difference between signals in two sets of 10 Cerenkov counters located at 
4.4 $<|\eta|<$ 4.9, to select collisions that were close to the nominal vertex position $z_{vtx}=0$ along 
the beam-axis. About 4 million events each of Cu+Cu and Au+Au collisions at $\sqrt{s_{_{NN}}}$ = 200~GeV were 
selected for further analysis by requiring that the primary collision vertex falls within $|z_{vtx}|<$ 6~cm.

The angular coordinates ($\eta,\phi$) of charged particles are measured using the location of the
energy deposited in the silicon pads of the Octagon. The granularity of the Octagon is determined 
by the sizes of the readout pads which are about $11.25^{\circ}$ ($\sim$ 0.2 radians) in $\phi$ and range 
from 0.006 to 0.05 in $\eta$. Noise and background hits 
are rejected by placing a lower threshold on the deposited energy corrected for the path length 
through the silicon after hit merging, assuming that the charged particle originated from the primary
vertex. Depending on $\eta$, merged hits with less than 50-60\% of the energy loss expected for a 
minimum ionizing particle are rejected. More details of the hit reconstruction procedure can be found in Ref.~\cite{hits_paper}.

\section{Analysis procedure}
\label{analysis}

\begin{figure*}[t!]
\vspace{-0.1cm}
\hspace{1.5cm}
\centerline{
  \mbox{\includegraphics[width=0.8\linewidth]{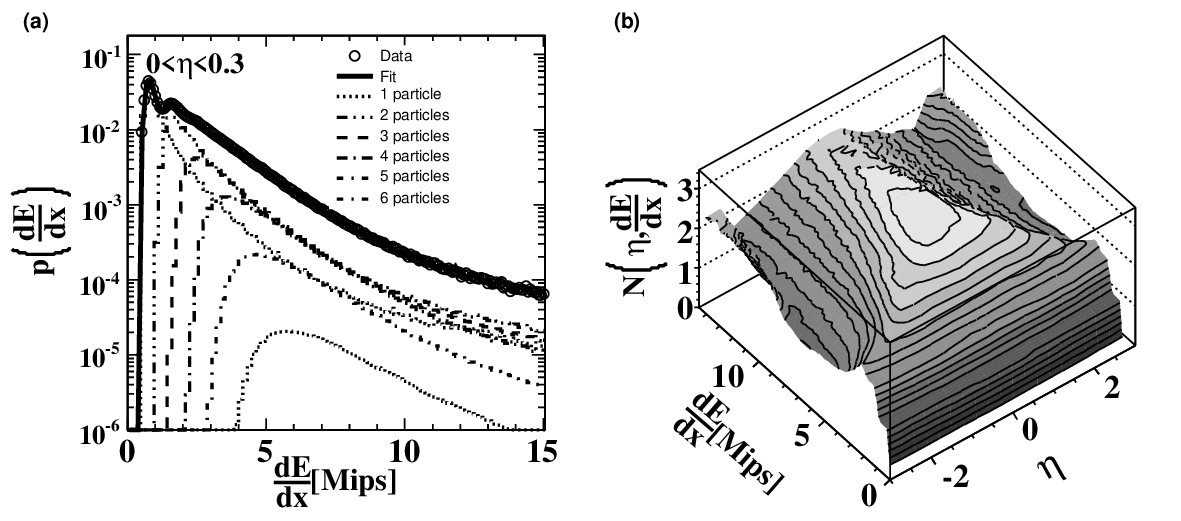}}}
\vspace{-0.3cm}
  \caption{ \label{SampledEdxfitForPaper_AuAuOn_binstart17_binnum1} 
                                 (a) $dE/dx$ distribution for hits registered in the Octagon in the 0 $<\eta<$ 0.3 range (open circles)
                                 fitted by a sum (solid line) of predicted $dE/dx$ distributions for various number
                                 of particles hitting a single pad (dashed lines) for the most central 3\% of 200~GeV 
                                 Au+Au collisions. A MIP is defined as the energy deposited by a single minimum-ionizing
                                 particle at normal incidence.
                                 (b) Estimated average number of particles per silicon pad as a function of $\eta$
                                 and $dE/dx$ for the most central 3\% of 200~GeV Au+Au collisions.
}
\vspace{-0.3cm}
\end{figure*}

The detailed analysis procedure is described in Ref.~\cite{phobos_pp}. The inclusive two-particle 
correlation function in ($\Delta \eta,\Delta \phi$) space is defined as follows:

\vspace{-0.4cm}
\begin{eqnarray}
\label{2pcorr_incl}
R(\Delta \eta,\Delta \phi)&=&\left<(n-1)\left(\frac{\rho^{\rm II} 
(\Delta \eta,\Delta \phi)}{\rho^{\rm mixed}(\Delta \eta,\Delta \phi)}-1\right)\right> \nonumber \\
&=&\frac{\left<(n-1)\rho^{\rm II}(\Delta \eta,\Delta \phi)\right>}{\rho^{\rm mixed}(\Delta \eta,\Delta \phi)}-\left<n-1\right>.
\end{eqnarray} 

\noindent The first line of Eq.~\ref{2pcorr_incl} shows the behavior of the correlation function, while
its mathematically equivalent second line corresponds to the analysis procedure described below.
For a given centrality bin, $\rho^{\rm II}(\Delta \eta,\Delta \phi)$ (normalized to unit integral) is the 
foreground pair distribution evaluated event-by-event by taking pairs within the same event. 
It is then weighted by a factor, $(n-1)$ (where $n$ is the total number of hits in each event, while generator level tracks
are used in MC), and averaged over all the events. The event multiplicity normalization factor, $(n-1)$, is introduced to compensate for the 
trivial dilution effects from uncorrelated particles since the number of uncorrelated pairs 
grows quadratically with $n$, while the number of correlated pairs grows only linearly. 
Therefore, $R(\Delta \eta,\Delta \phi)$ is defined in such a way that if a heavy ion 
collision is simply a superposition of individual p+p collisions, the same correlation 
function will be observed in both A+A and p+p collisions.

The mixed-event background distribution 
$\rho^{\rm mixed}(\Delta \eta,\Delta \phi)$ (also normalized to unit integral) is constructed 
by randomly selecting two particles from two different events with similar vertex (with a bin width of 0.2~cm)
and centrality (in bins of 5\% of the total inelastic cross section), representing a product of two single particle distributions. 
Since the background is found to be roughly multiplicity independent within a centrality bin, the inclusive
$\rho^{\rm mixed}(\Delta \eta,\Delta \phi)$ is used in our calculations of Eq.~\ref{2pcorr_incl} for each centrality class.
Thus it can be either inside or outside the event average as shown in the two lines of Eq.~\ref{2pcorr_incl}. 
$R(\Delta \eta,\Delta \phi)$ is symmetrized at $\Delta\eta$=0 and $\Delta\phi$=0 following the same convention as 
in the previous publication~\cite{phobos_pp}.
As can be inferred from the first line of Eq.~\ref{2pcorr_incl}, since both
$\rho^{\rm II}(\Delta \eta,\Delta \phi)$ and $\rho^{\rm mixed}(\Delta \eta,\Delta \phi)$ are normalized to
unit integral, $R(\Delta \eta,\Delta \phi)$ will range from positive to negative. 
Therefore, there is no particular significance when $R(\Delta \eta,\Delta \phi)$=0 in a small region of 
$\Delta \eta$ and $\Delta \phi$, although $R(\Delta \eta,\Delta \phi)$=0 everywhere would indicate the absence of correlations.

Corrections for secondary effects and incomplete acceptance have 
been applied in a similar way to those in p+p collisions resulting in corrections with comparable magnitudes~\cite{phobos_pp}. 
Since the PHOBOS Octagon is a single-layer silicon detector, there is no $p_{T}$, charge 
or mass information available for the particles. All charged particles above a low-$p_{T}$ 
cutoff of about 7~MeV/c at $\eta$=3, and 35~MeV/c at $\eta$=0 (which is the threshold below which a 
charged particle is stopped by the beryllium beam pipe) are included on equal footing.
Thus secondary effects, such as $\delta$-electrons, $\gamma$ conversions and weak decays, cannot be all rejected directly. 
The incomplete azimuthal acceptance in some pseudorapidity regions naturally suppresses the overall 
correlation strength, but MC simulations show that it does not change the shape of the correlation 
function $R(\Delta \eta, \Delta \phi)$. To correct for these detector effects in the data,
correlation functions are calculated for MC events (PYTHIA for p+p collisions and HIJING for A+A collisions) at $\sqrt{s}$ = 200~GeV
both at the generator level for true primary charged hadrons, $R_{\rm pri}^{\rm MC}(\Delta \eta, \Delta \phi)$, 
and with the full GEANT detector simulation and reconstruction procedure, $R_{\rm sim}^{\rm MC}(\Delta \eta, \Delta \phi)$.
The whole correction procedure can be summarized by the following equation:

\vspace{-0.4cm}
\begin{equation}
\label{corrected_corr_scale_corrected}
R_{\rm final}^{\rm data}(\Delta \eta, \Delta \phi) = A \times [R_{\rm raw}^{\rm data}
(\Delta \eta, \Delta \phi)-S(\Delta \eta, \Delta \phi)]. 
\end{equation}

\noindent where the function,

\vspace{-0.4cm}
\begin{equation}
\label{secondary_corr}
S(\Delta \eta, \Delta \phi) = R_{\rm sim}^{\rm MC}(\Delta \eta, \Delta \phi)
-R_{\rm pri,acc}^{\rm MC}(\Delta \eta, \Delta \phi)
\end{equation}

\noindent represents the correction for secondary effects which mainly contribute to 
a large, narrow spike in $R(\Delta \eta, \Delta \phi)$ around $\Delta \eta$=0, 
$\Delta \phi$=0 (for more details, see Ref.~\cite{phobos_pp}). The acceptance correction factor, $A$, is estimated to 
be around 1.3 with little dependence on vertex and centrality. The total systematic uncertainty
for these corrections is typically about 5\% of the final value of $R(\Delta \eta,\Delta \phi)$.

In addition to these corrections, the high occupancies measured in A+A collisions (which are in the range from 50-60\% at midrapidity for the most 
central 3\% of 200~GeV Au+Au collisions) require us to account 
for the high probability of multiple particles hitting a single pad. The $dE/dx$ distribution 
of hits in a very low multiplicity environment (e.g. 55-60\% peripheral Cu+Cu which has an 
occupancy of about 4\% at midrapidity) has been measured first in a narrow $\eta$ bin of 
0.3 unit (same bin size as for $R(\Delta \eta,\Delta \phi)$), to approximate the $dE/dx$ distribution for a single 
particle hitting a single silicon pad, $p_1(\frac{dE}{dx})$. By convoluting $p_1(\frac{dE}{dx})$ 
$i$ times, the $dE/dx$ distribution of $i$ particles hitting a single pad can be predicted, 
$p_i(\frac{dE}{dx})$ (i=1, 2, 3....). Then, the $dE/dx$ distribution in more central 
data events, $p(\frac{dE}{dx})$, is fitted by a sum of $p_i(\frac{dE}{dx})$ 
with weighting factor $w_i$ such that the relative contribution of different number 
of particles hitting on a single pad can be estimated, as illustrated in 
Fig.~\ref{SampledEdxfitForPaper_AuAuOn_binstart17_binnum1}a for the most central 3\% of 200~GeV 
Au+Au collisions as an example. While the $w_i$ are free parameters in these fits, they have been compared 
to a Poisson distribution and found to be in very good agreement.
The average number of particles per pad can thus be calculated as a function of $dE/dx$, 

\vspace{-0.4cm}
\begin{equation}
\label{n_dEdx}
N\left(\frac{dE}{dx}\right)=\frac{\displaystyle\sum_i i \times w_ip_i\left(\frac{dE}{dx}\right)}{\displaystyle\sum_i w_ip_i\left(\frac{dE}{dx}\right)}.
\end{equation}

\noindent Performing the procedure above in each $\eta$ and centrality bin, the average number of particles per pad as 
a function of $dE/dx$ and $\eta$, $N(\eta,\frac{dE}{dx})$, is derived and shown in 
Fig.~\ref{SampledEdxfitForPaper_AuAuOn_binstart17_binnum1}b for the most central 3\% of 200~GeV 
Au+Au collisions. In calculating the correlation function, each hit is assigned a 
weight $N(\eta,\frac{dE}{dx})$ based on its $\eta$ and $dE/dx$. Using this procedure, the
effects of high occupancy are compensated at the hit level. The systematic uncertainty of the occupancy correction
is about 5\%-7\% ranging from peripheral to central collisions. The effect of occupancy fluctuations within 
a particular centrality bin has also been investigated. The analysis was repeated both for a much narrower range in multiplicity 
in the same bin, as well as using the correction weights for a higher-multiplicity bin, and the impact on 
the final results has been found to be negligible.

\begin{figure*}[thb]
\vspace{-0.5cm}
\hspace{1.5cm}
\centerline{
  \mbox{\includegraphics[width=0.9\linewidth]{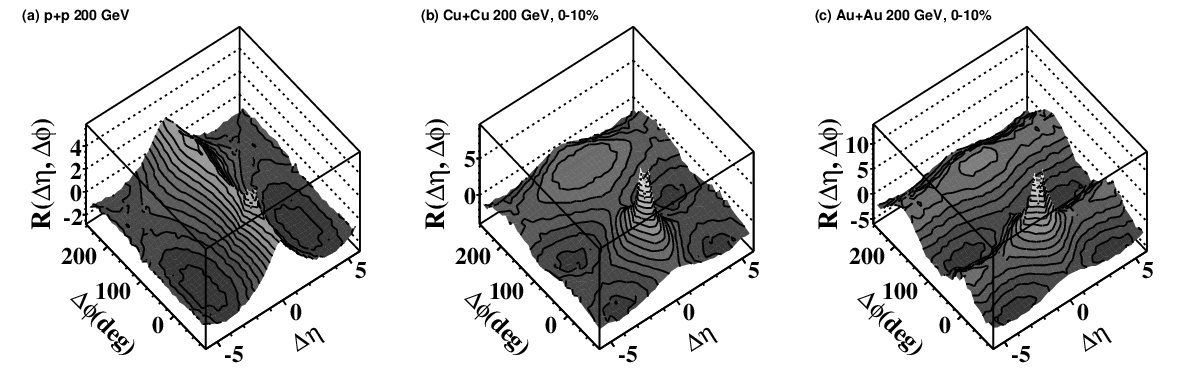}}}
\vspace{-0.3cm}
  \caption{ \label{EtaPhiCFFor3} 
                             Two-particle angular correlation functions in $\Delta \eta$ and $\Delta \phi$
                             for (a) p+p, the most central 10\% (b) Cu+Cu and (c) Au+Au 
                             collisions at \mbox{$\sqrt{s}$} or \mbox{$\sqrt{s_{_{NN}}}$ = 200~GeV}.
}
\end{figure*}

\begin{figure*}[thb]
\vspace{-0.1cm}
\hspace{1.5cm}
\centerline{
  \mbox{\includegraphics[width=0.9\linewidth]{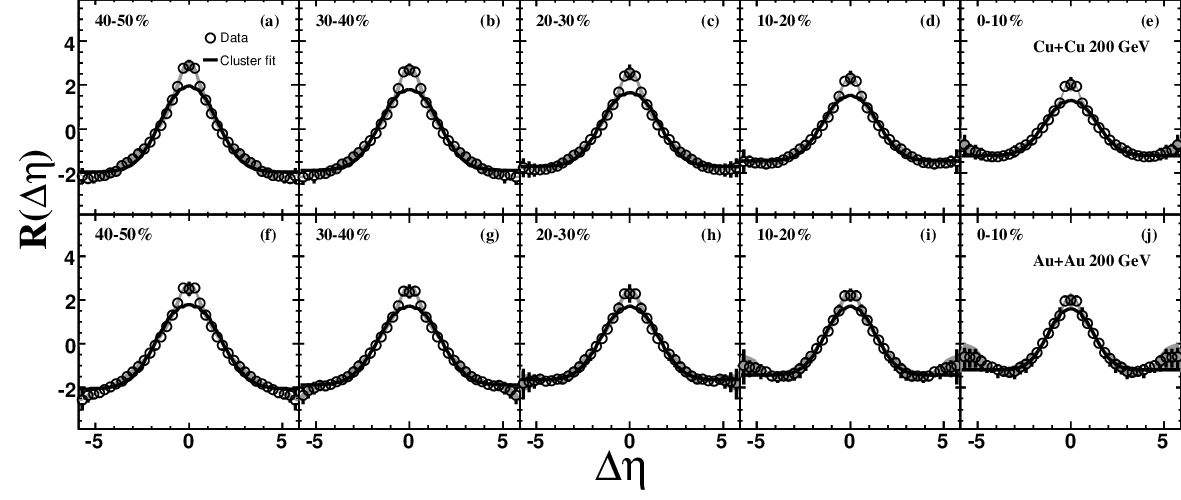}}}
\vspace{-0.3cm}
  \caption{ \label{EtaCFAll}  
  Two-particle pseudorapidity correlation functions, averaged over the $\Delta \phi$ 
  range from $0^{\circ}$ to $180^{\circ}$, in Cu+Cu (upper row) and Au+Au (lower row) 
  collisions for five different centrality classes at \mbox{$\sqrt{s_{_{NN}}}$ = 200~GeV}.
  The solid curves (black) correspond to Eq.~\ref{2pcorr_clusterfitting_incl} with the final 
  values of the parameters (see text for discussion). 
  The error bars and bands (grey) correspond to point-to-point systematic errors 
  and overall scale errors respectively with 90\% C.L. The statistical errors are negligible.
}
\vspace{-0.3cm}
\end{figure*}

\section{Results}
\label{results}

The final 2-D two-particle inclusive correlation functions for charged particles after
all corrections are shown in Fig.~\ref{EtaPhiCFFor3} as a function of $\Delta \eta$ and 
$\Delta \phi$ for the most central 10\% Cu+Cu (b) and Au+Au (c) collisions at $\sqrt{s_{_{NN}}}$ = 200~GeV. 
For comparison, the previous measurement in p+p collisions (a) at the same energy~\cite{phobos_pp} is also shown. 
In p+p collisions, the complex 2-D correlation structure is approximately Gaussian in $\Delta\eta$
and persists over the full $\Delta\phi$ range, becoming broader toward larger $\Delta\phi$.
This feature can be qualitatively described by an independent cluster approach, as will be shown later.
Compared to p+p collisions, heavy ion collisions show not only the cluster-like structure, but also 
a $\cos(2\Delta\phi)$ modulation due to elliptic flow~\cite{CF_flow}.  

As was done in the analysis of p+p collisions~\cite{phobos_pp}, the 2-D correlation functions 
in A+A are integrated over $\Delta\phi$ to give the 1-D correlation 
functions $R(\Delta\eta)$, shown in Fig.~\ref{EtaCFAll}. This allows a quantitative study of cluster
properties in pseudorapidity space, with the elliptic flow contribution averaging to zero.
$R(\Delta\eta)$ is then fitted to a functional form derived in Ref.~\cite{cluster_fit} 
in an independent cluster emission model:

\begin{equation}
\label{2pcorr_clusterfitting_incl}
R(\Delta \eta)=\alpha\left[\frac{\Gamma(\Delta \eta)}{\rho^{\rm mixed}(\Delta \eta)}-1\right]   
\end{equation}

\noindent where $\rho^{\rm mixed}(\Delta \eta)$ is the background distribution obtained by event-mixing
and averaged over $\Delta\phi$. The parameter $\alpha$ is equal to $K_{\rm eff}-1$, where $K_{\rm eff}$
is the effective cluster size, defined as:

\begin{equation}
\label{keff} 
K_{\rm eff}=\alpha + 1=\langle K \rangle+\frac{\sigma_{K}^{2}}{\langle K \rangle}
\end{equation}

\noindent which depends on the first two moments (mean and sigma) of the distribution of 
cluster size $K$ (number of particles decayed from a cluster) over all clusters in all events. The function 
$\Gamma(\Delta \eta)$ is a normalized Gaussian function $\frac{1}{\sqrt{4\pi}\delta} exp{(-(\Delta \eta)^{2}/(4\delta^{2}))}$, 
where the $\delta$ parameter is equal to the width of the $\Delta\eta$ distribution of the particle pairs from
a single cluster. It is connected with another variable also characterizing the cluster width, 
$\sigma_{\eta}$ (understood as the width of the distribution of the difference $\eta_{\rm particle}-\eta_{\rm cluster}$), 
by the formula: $\delta = \sqrt{\frac{K}{K-1}} \times \sigma_{\eta} $, for fixed $K$. The factor $\sqrt{\frac{K}{K-1}}$
difference is due to the fact that the average of $\eta_{\rm particle}$ in a cluster 
is constrained to be conserved (equal to $\eta_{\rm cluster}$).
Of course, without direct knowledge of the distribution of $K$, the average cluster size $\langle K \rangle$ 
and width $\sigma_{\eta}$ cannot be derived based on $K_{\rm eff}$ and $\delta$. 

Correlation functions for bins in vertex and centrality 
are individually fit using Eq.~\ref{2pcorr_clusterfitting_incl} to extract the effective 
cluster size, $K_{\rm eff}$, and the cluster decay width, $\delta$, for each bin. These results are then 
averaged over the vertex range to find the final results for each centrality. The averaged 
correlation functions are shown in Fig.~\ref{EtaCFAll} along with a line showing 
Eq.~\ref{2pcorr_clusterfitting_incl} with the final averaged values of the fit parameters. 
The three most central points (a region of $|\Delta\eta|<$ 0.45) in $R(\Delta\eta)$ are 
excluded from the fits mainly due to the large uncertainty stemming from residual detector effects. 

\begin{figure}[ht]
\hspace{-0.3cm}
\centerline{
  \mbox{\includegraphics[width=0.8\linewidth]{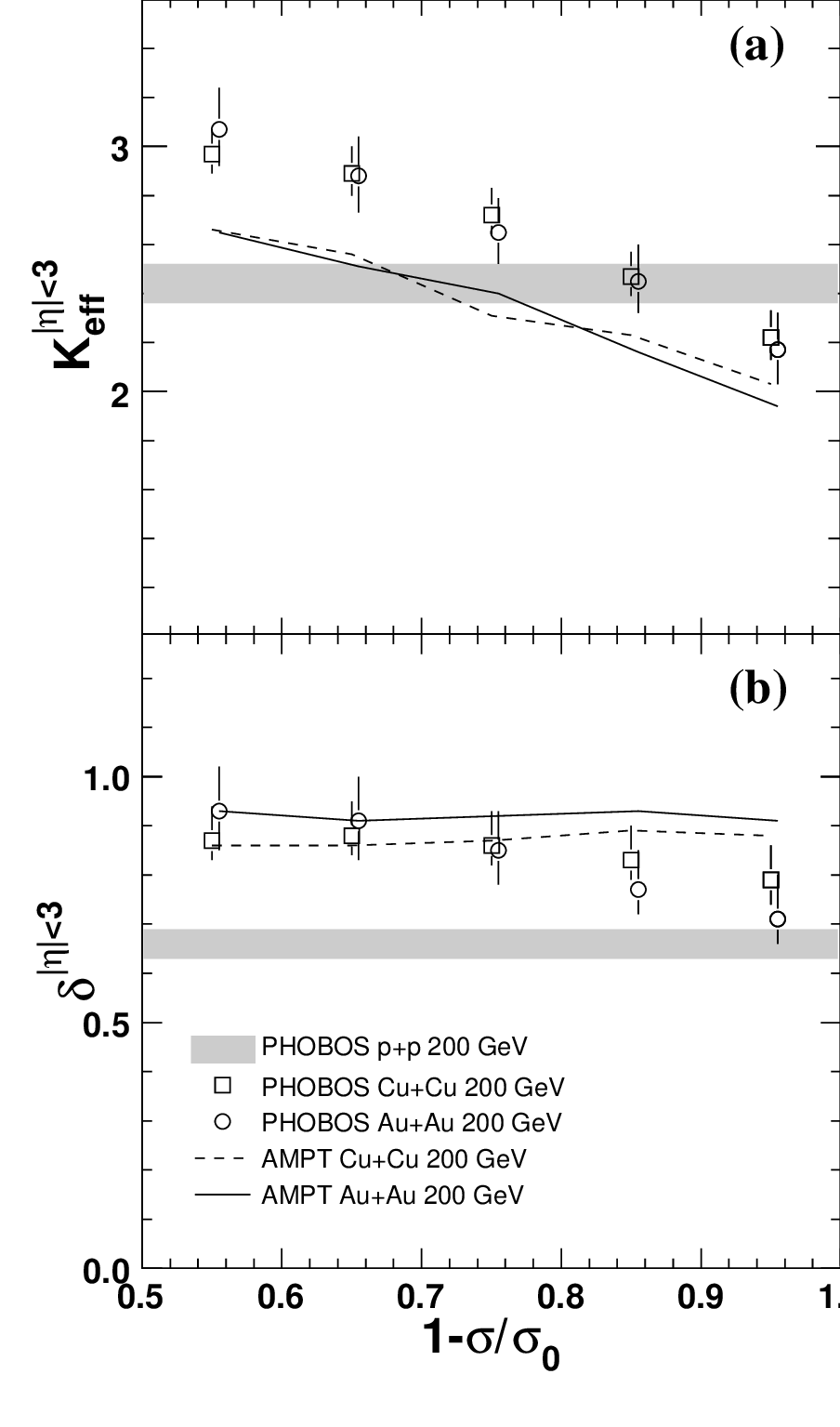}}}
\vspace{-0.3cm}
  \caption{
            (a) $K_{\rm eff}$ and (b) $\delta$ as a 
            function of fractional cross section (1-$\sigma$/$\sigma_{0}$=1 for the most central collisions) for PHOBOS data (open symbols) 
            and from the AMPT model (lines) in Cu+Cu (squares) and Au+Au 
            (circles) collisions for $|\eta|<$ 3 at \mbox{$\sqrt{s_{_{NN}}}$ = 200~GeV}. 
            The error bars for data points represent systematic errors with 90\% C.L.
            Results from p+p collisions at \mbox{$\sqrt{s}$ = 200~GeV}~\cite{phobos_pp} 
            are shown by the shaded band.
          }
  \label{clustervscross_incl}
\vspace{-0.3cm}
\end{figure}

\begin{figure}[ht]
\hspace{-0.3cm}
\centerline{
  \mbox{\includegraphics[width=\linewidth]{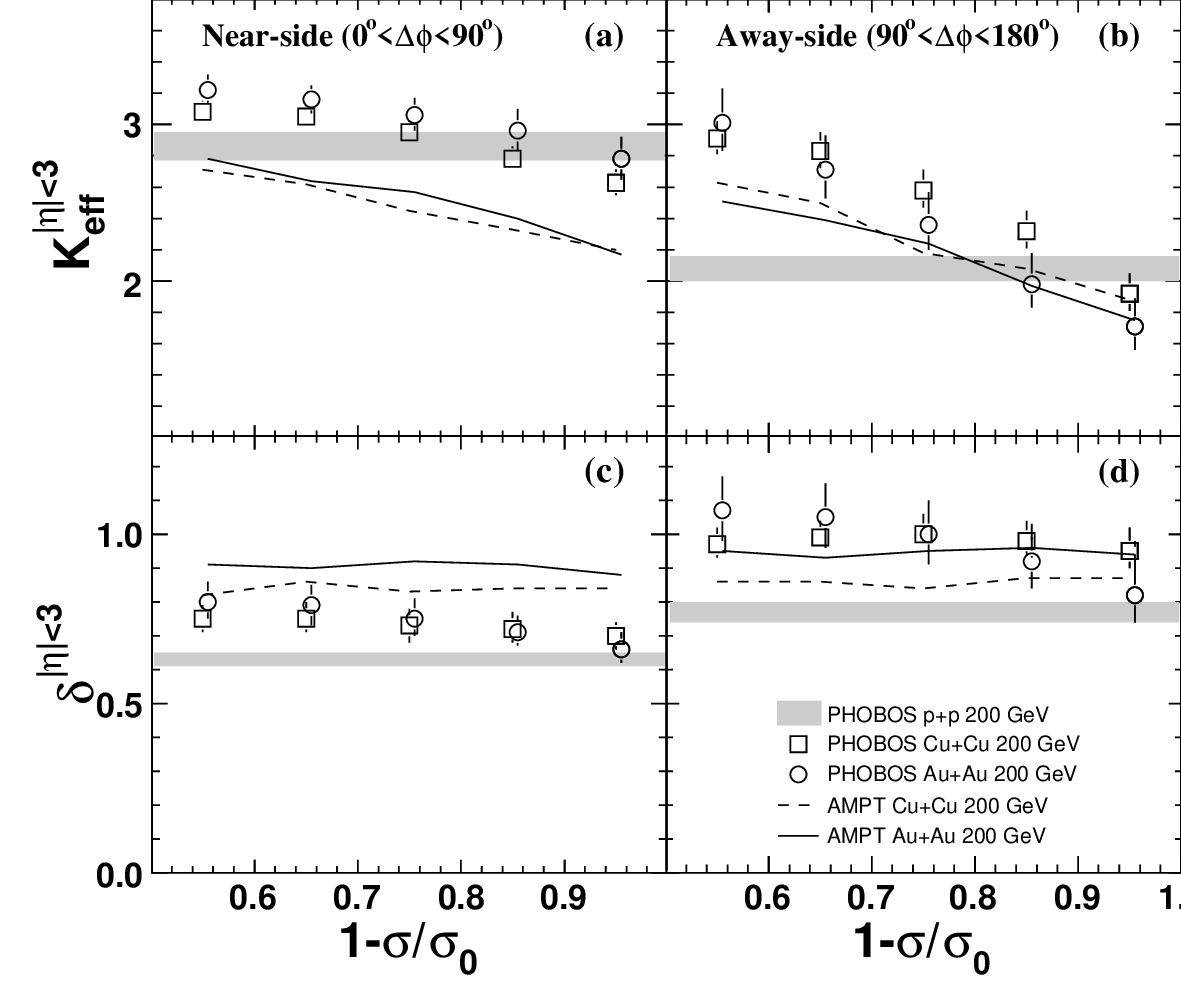}}}
\vspace{-0.3cm}
  \caption{
            Near-side (left column) and away-side (right column) $K_{\rm eff}$ 
            (upper row) and $\delta$ (lower row) as a function of fractional 
            cross section (1-$\sigma$/$\sigma_{0}$=1 for the most central collisions) for PHOBOS data (open symbols) 
            and from the AMPT model (lines) in Cu+Cu (squares) and Au+Au 
            (circles) collisions for $|\eta|<$ 3
            at \mbox{$\sqrt{s_{_{NN}}}$ = 200~GeV}. The error bars for data points 
            represent systematic errors with 90\% C.L. Results from p+p collisions 
            at \mbox{$\sqrt{s}$ = 200~GeV}~\cite{phobos_pp} are shown by the shaded band.
          }
  \label{clustervscross_nearaway}
\vspace{-0.3cm}
\end{figure}

Results on effective cluster size ($K_{\rm eff}$) and decay width 
($\delta$) as a function of the fractional cross section $1-\sigma/\sigma_{0}$, 
where $\sigma_{0}$ is the total A+A inelastic cross section, 
are shown in Fig.~\ref{clustervscross_incl} for Cu+Cu 
and Au+Au collisions at $\sqrt{s_{_{NN}}}$ = 200~GeV. The systematic uncertainties are estimated
using a similar procedure to p+p collisions~\cite{phobos_pp} with an additional contribution from 
the occupancy corrections. The overall scale error, common to both Cu+Cu and Au+Au, 
is 5\% for both $K_{\rm eff}$ and $\delta$.
The shaded band indicates the value found in $\sqrt{s}$ = 200~GeV p+p collisions, which suggests 
that the cluster properties are similar in p+p and A+A systems. This implies that the 
phenomenological properties of hadronization appear to be similar in p+p and A+A. 
However, an increase of both the effective cluster size and decay width is 
observed going from p+p to peripheral A+A systems. Toward more central collisions, 
it is also observed that the effective cluster size systematically decreases with increasing collision 
centrality in both Cu+Cu and Au+Au collisions, whereas the cluster
decay width is approximately constant over the whole centrality range within the systematic 
uncertainties. Furthermore, by comparing the two systems at the same fraction of the inelastic 
cross section (which is related to the ratio of impact parameter $b$ to 
the radius $R$ of the nucleus, $b/2R$), a ``geometric scaling'' feature is revealed, which shows 
a similar effective cluster size at the same collision geometry of the system, i.e. the shape of the overlap region.
This feature is not obviously expected as the cluster parameters are constructed to reflect short-range correlations in rapidity and thus are not directly connected with the overall geometry of the initial state of the collision.
Comparison of the data with AMPT~\cite{AMPT} shows that the model
gives the same qualitative trend as the data in the same $\eta$ acceptance, but with $K_{\rm eff}$ 
values systematically lower by about 0.4.
Note that the values of $K_{\rm eff}$ and $\delta$ are 
extracted in a limited acceptance of $|\eta|<$ 3, and therefore
are normally smaller than for a full acceptance measurement. The acceptance effect will 
be discussed quantitatively in the next section. In AMPT at the generator level, the decrease in effective cluster size 
with increasing event centrality appears to be related to the hadronic rescattering stage. 
Turning off hadronic rescattering processes in AMPT leads to a larger effective cluster size in both Au+Au and 
Cu+Cu that is approximately invariant for all centralities.

Further detailed studies on cluster properties have also been performed. The entire analysis was repeated
for pairs in the restricted $\Delta\phi$ range. Instead of averaging over
the whole $\Delta\phi$ region, the cluster parameters can be extracted for pairs in the near-side and away-side 
$\Delta\phi$ ranges ($0^{\circ}<\Delta\phi<90^{\circ}$ and $90^{\circ}<\Delta\phi<180^{\circ}$ 
respectively). Clusters with high $p_{T}$ generally contribute more pairs at the near-side, while away-side
pairs mainly come from lower $p_{T}$ clusters. In this restricted averaging, the $\cos(2\Delta\phi)$ 
elliptic flow component again averages to zero.
The results are shown in Fig.~\ref{clustervscross_nearaway} as a function of fractional cross section
for Cu+Cu and Au+Au collisions at $\sqrt{s_{_{NN}}}$ = 200~GeV. The overall scale error is 3\% for the near-side data and 5\%
for the away-side data, for both $K_{\rm eff}$ and $\delta$. Over the studied centrality range,
the effective cluster size extracted for pairs from away-side decreases by about 30-40\% with increasing centrality, whereas 
the decrease for the near-side is somewhat smaller. Such a behavior could be understood in a scenario 
where the medium is extremely dense at more central collisions and 
only clusters produced close to the surface can survive. Then, for low $p_{T}$ clusters, it is more 
likely that part of its decay particles travel into the medium and get absorbed, resulting in 
a suppression of away-side correlations. As for the observed collision geometry scaling of the effective cluster size, 
it might be related to the surface to volume ratio of the system. More detailed modeling is still 
being investigated to understand these phenomena. In this case, AMPT shows a smaller difference of cluster properties
between near and away-side than that observed in the data.

\section{Correcting for limited acceptance}
\label{acceptance}

As mentioned in the previous section, some particles from cluster decay fall outside of the 
PHOBOS Octagon detector acceptance in pseudorapidity ($|\eta|<$ 3). This reduces both $K_{\rm eff}$ and $\delta$.
In order to quantitatively study this effect, a simple independent cluster model (ICM) as well as several 
dynamical models are used. In our ICM approach, for each event clusters are 
generated, each with a given mass, transverse momentum, $p_{T}$, and $\eta$.
The cluster decays isotropically in its rest frame into $K$ particles (assumed to be pions) 
constrained by the available phase space. The mass of the clusters is taken to be 
$0.35 \times K$~(GeV/$c^{2}$). The $p_{T}$ and $\eta$ of the clusters are drawn from 
distributions that have been tuned such that the final inclusive charged particles match 
the measured spectra~\cite{phobos_dNdeta_AuAu,phobos_dNdeta_CuCu,star_meanpT}. 
In addition, the global momentum of the clusters is always conserved event-by-event in order to 
preserve the $\cos(\Delta\phi)$ component typically seen in the p+p 2-D correlation function 
(e.g. in Fig.~\ref{EtaPhiCFFor3}a).

\begin{figure}[b!]
\vspace{-0.2cm}
\hspace{1.5cm}
\centerline{
  \mbox{\includegraphics[width=\linewidth]{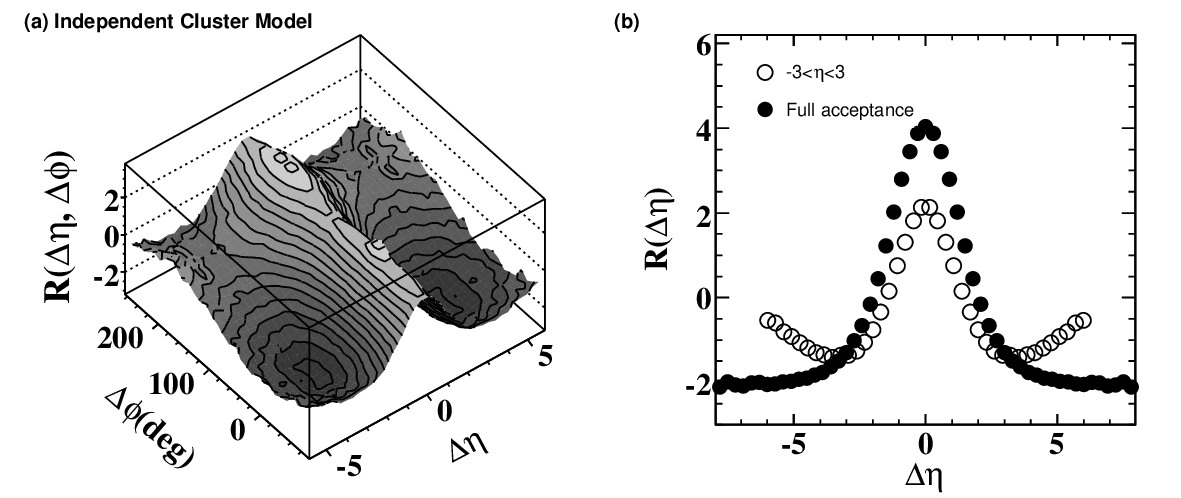}}}
\vspace{-0.3cm}
  \caption{ \label{Paper_icm_mod33_comparison}
            (a) 2-D two-particle correlation function in $\Delta \eta$ and 
            $\Delta \phi$ for ICM with $K=3$ and $\gamma=0$ (defined in Eq.~\ref{clusterwidth_A}).
            (b) Comparison of 1-D pseudorapidity correlation function for 
            ICM with $K=3$ and $\gamma=0$ (defined in Eq.~\ref{clusterwidth_A}) between the acceptance of $|\eta|<$ 3 and the full acceptance.
}
\vspace{-0.1cm}
\end{figure}

\begin{figure}[thb]
\vspace{-0.2cm}
\hspace{1.5cm}
\centerline{
  \mbox{\includegraphics[width=\linewidth]{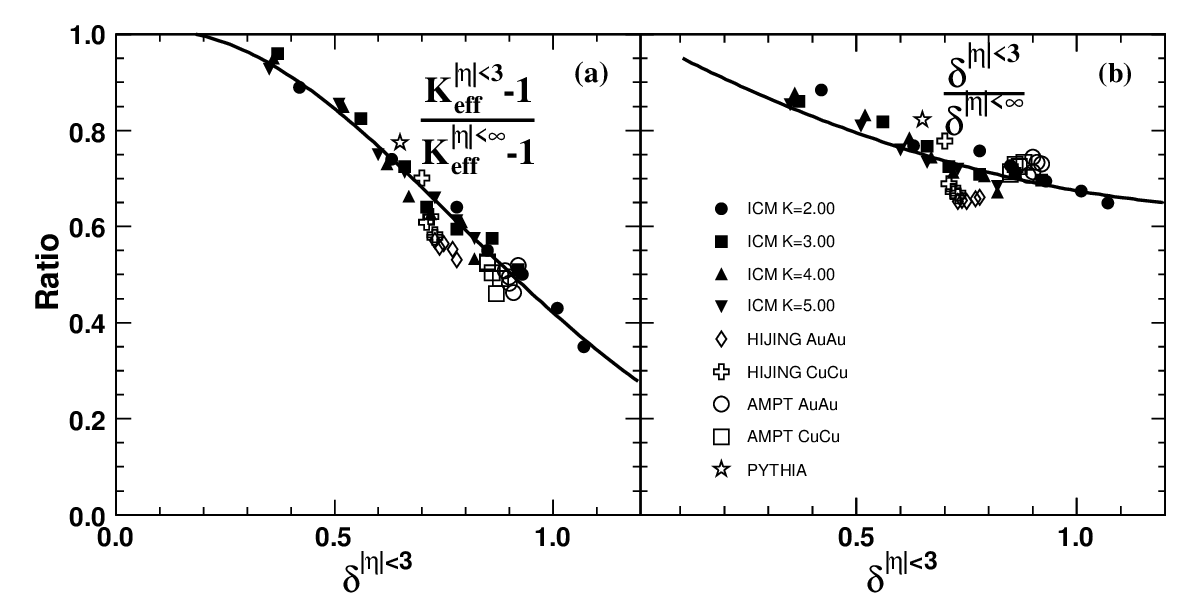}}}
\vspace{-0.3cm}
  \caption{ \label{ICM_d3vsdmax_fit_new_binobs2}
            (a) Ratio of $K^{\rm |\eta|<3}_{\rm eff}$-1 to $K^{\rm |\eta|<\infty}_{\rm eff}$-1
            and (b) $\delta^{\rm |\eta|<3}$ to $\delta^{\rm |\eta|<\infty}$ as a function of 
            $\delta^{\rm |\eta|<3}$ obtained in ICM (solid symbols) as well as PYTHIA, 
            HIJING and AMPT models (open symbols). The solid line is a smooth function fit to all the
            models.
}
\vspace{-0.2cm}
\end{figure}

\begin{figure*}[thb]
\vspace{-0.5cm}
\hspace{1.5cm}
\centerline{
  \mbox{\includegraphics[width=0.85\linewidth]{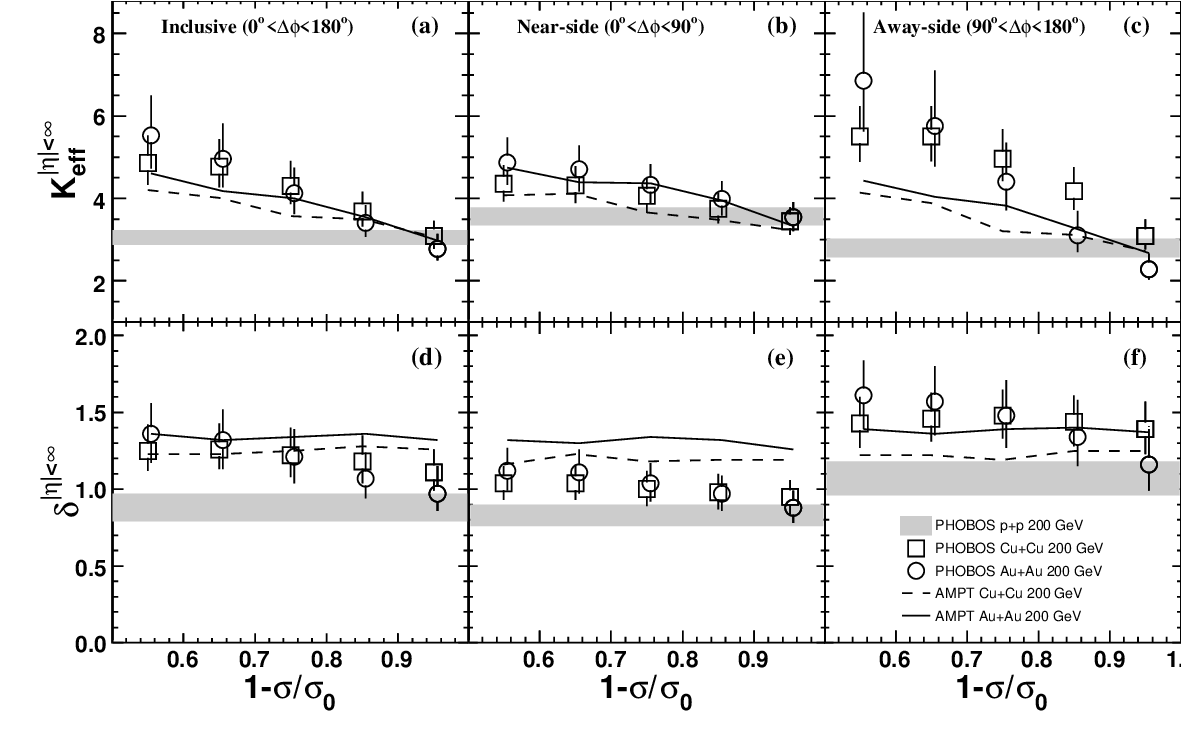}}}
\vspace{-0.3cm}
  \caption{ \label{clustervscross_nearaway_corr}
            Inclusive (left column), near-side (middle column) and away-side (right column) 
            $K_{\rm eff}$ (upper row) and $\delta$ (lower row)  
            as a function of fractional cross section (1-$\sigma$/$\sigma_{0}$=1 for the most central collisions) measured by PHOBOS (open symbols) 
            and from the AMPT model (lines) after correcting to the full acceptance in 
            Cu+Cu (squares) and Au+Au (circles) collisions at \mbox{$\sqrt{s_{_{NN}}}$ = 200~GeV}. 
            The error bars for data points represent systematic errors with 90\% C.L.
            Results in p+p collisions at \mbox{$\sqrt{s}$ = 200~GeV} after correcting to the full acceptance 
            are shown in the shaded band, which are larger than those in Ref.~\cite{phobos_pp} 
            measured in an acceptance of $|\eta|<$ 3.
}
\vspace{-0.3cm}
\end{figure*}

In Fig.~\ref{Paper_icm_mod33_comparison}a, an example of a 2-D two-particle correlation 
function from ICM is shown with $K=3$ measured in an acceptance of $|\eta|<$ 3. It shows a 
qualitatively similar structure to what is observed in p+p collisions in Fig.~\ref{EtaPhiCFFor3}a, a Gaussian
shape short-range correlation along $\Delta\eta$ which gets wider going from near-side to away-side.
The narrow peak in the near-side part of the correlation function is primarily from higher 
$p_{T}$ clusters as well as 3-body decays (e.g. in Ref.~\cite{ISR_twolowenergy}), whereas the broader away-side 
generally arises from the decay of clusters with lower $p_{T}$.
Fig.~\ref{Paper_icm_mod33_comparison}b shows a comparison between 1-D $\Delta\eta$ correlation
functions measured in the limited acceptance of $|\eta|<$ 3 and the full acceptance. 
From this comparison, one can see that the shape is significantly modified by the limited acceptance, resulting in reductions in both $K_{\rm eff}$ and $\delta$. By comparing the cluster parameters extracted for the 
acceptance of $|\eta|<$ 3, $K^{\rm |\eta|<3}_{\rm eff}$ and $\delta^{\rm |\eta|<3}$, with 
the full acceptance, $K^{\rm |\eta|<\infty}_{\rm eff}$ and $\delta^{\rm |\eta|<\infty}$
an acceptance correction can be obtained in the context of the ICM. 

Note that in the ICM, $\delta$ 
is not an independent variable, but rather depends on the cluster $p_T$ for given cluster mass. 
The higher the $p_T$ is, the narrower the cluster width will be. In a scenario of isotropic decay of the clusters, 
the maximum possible width turns out to correspond to $\delta^{\rm |\eta|<3} \sim 0.75$ for $K=2$ and $p_T$ 
of all clusters fixed at 0, which is lower than the cluster width observed in A+A collisions 
($\sim 0.9$ in the most peripheral collisions). Therefore, to generate a wider range of possible $\delta$ 
parameter values in the ICM, another parameter, $\gamma$, is introduced to artificially modify the width of the clusters 
by rescaling the relative pseudorapidity of each decay product:
\begin{equation}
\label{clusterwidth_A}
\eta^{\prime}_{i}-\eta_{0} = \gamma \times(\eta_{i}-\eta_{0}), i=1,2,3...
\end{equation}
\noindent where $\eta_{0}$ represents the pseudorapidity of the original clusters, and $\eta_{i}$ and $\eta^{\prime}_{i}$
correspond to the pseudorapidity of decaying particles from clusters before and after modification.
In this way, any value of the cluster width can be obtained while keeping the original $\eta$ of the cluster unchanged.
For each set of $K$ and $\gamma$, the $dN/d\eta$ and 
$dN/dp_{T}$ distributions of the clusters are tuned to match those of the final state inclusive 
charged particles measured in the data.

The ICM, with a range of values of $K$ and $\gamma$, has been used to generate a set of two-particle correlation
functions, both for $|\eta|<$ 3 and the full acceptance. In Fig.~\ref{ICM_d3vsdmax_fit_new_binobs2}, the ratios
($K^{\rm |\eta|<3}_{\rm eff}$-1)/($K^{\rm |\eta|<\infty}_{\rm eff}$-1) and $\delta^{\rm |\eta|<3}$/$\delta^{\rm |\eta|<\infty}$ from the ICM are shown as a function of $\delta^{\rm |\eta|<3}$, as extracted directly from fits to the correlation
function. It is clearly seen that the suppression of these ratios are primarily a function of $\delta^{\rm |\eta|<3}$ 
only, and both $K_{\rm eff}$ and $\delta$ are suppressed more as $\delta^{\rm |\eta|<3}$ increases. This is because 
correlated particles are more likely to fall outside the measured region as the cluster width increases. 
The suppression factors in the dynamical models like PYTHIA~\cite{PYTHIA} for p+p, HIJING~\cite{HIJING}, AMPT for 
Cu+Cu and Au+Au at various 
centralities are also calculated. All models are consistent within about 5\%-10\%. A second order polynomial function
is fitted to the values of all models in order to generate a smooth correction function. This function is 
applied to the measured $K^{\rm |\eta|<3}_{\rm eff}$ and $\delta^{\rm |\eta|<3}$ 
using Eq.~\ref{acceptance_correction} in order to estimate $K^{\rm |\eta|<\infty}_{\rm eff}$ 
and $\delta^{\rm |\eta|<\infty}$ for the experimental data:

\begin{eqnarray}
\label{acceptance_correction}
(K^{\rm |\eta|<\infty}_{\rm eff}-1)_{\rm data} &=& \frac{(K^{\rm |\eta|<\infty}_{\rm eff}-1)_{\rm MC}}{(K^{\rm |\eta|<3}_{\rm eff}-1)_{\rm MC}} \times (K^{\rm |\eta|<3}_{\rm eff}-1)_{\rm data} \nonumber \\
\delta^{\rm |\eta|<\infty}_{\rm data} &=& \frac{\delta^{\rm |\eta|<\infty}_{\rm MC}}{\delta^{\rm |\eta|<3}_{\rm MC}} \times \delta^{\rm |\eta|<3}_{\rm data}
\end{eqnarray}
 
The scattering of the points around the fitted correction curves in Fig.~\ref{ICM_d3vsdmax_fit_new_binobs2} 
is taken into account as one source of the systematic uncertainties on the acceptance correction procedure. 
As a cross check, the cluster parameters in the data have also been measured in $|\eta|<$ 2. 
The ratios, ($K^{\rm |\eta|<2}_{\rm eff}-1$)/($K^{\rm |\eta|<3}_{\rm eff}-1$) and 
$\delta^{\rm |\eta|<2}$/$\delta^{\rm |\eta|<3}$, are found to be consistent with the ICM and dynamical models.
The residual discrepancies between the results extrapolated to full phase space from $|\eta|<$ 2 and $|\eta|<$ 3
are used to estimate a separate contribution to the systematic uncertainty on the acceptance correction. 
The total uncertainty on the correction is thus found to be 12\% for $K_{\rm eff}$ and 9\% for $\delta$.

After applying the acceptance correction, the cluster parameters at full phase space, 
$K^{\rm |\eta|<\infty}_{\rm eff}$ and $\delta^{\rm |\eta|<\infty}$, in p+p, Cu+Cu and Au+Au collisions at 
$\sqrt{s_{_{NN}}}$ = 200~GeV are shown in Fig.~\ref{clustervscross_nearaway_corr}
for inclusive (left column), near-side (middle column) and away-side (right column) along with the results from AMPT. 
The systematic errors come not only from the measurement itself but also the acceptance correction procedure.
The values of $K_{\rm eff}$ and $\delta$ in p+p collisions extrapolated to full phase space are larger
than those presented in Ref.~\cite{phobos_pp} measured in a limited acceptance of $|\eta|<$ 3,
and better reflect the properties of the clusters produced in these reactions.
Since $\delta$ measured in both Cu+Cu and Au+Au collisions at PHOBOS only weakly
depends on centrality, the geometric scaling feature of $K_{\rm eff}$ between the two systems 
still holds after the acceptance correction as shown in Fig.~\ref{clustervscross_nearaway_corr}.
That said, the large values of $K^{\rm |\eta|<\infty}_{\rm eff}$ and $\delta^{\rm |\eta|<\infty}$ clearly pose a
challenging question as to the origin of such strong correlations with such a long range. 
In calculations from the Therminator model that include all known resonances~\cite{Therminator}, $K_{\rm eff}$ is 
approximately 2 and $\delta$ is no larger than 0.75, while in peripheral A+A collisions, there 
appear to be clusters that decay into 5-6 charged particles with much 
larger $\delta$. This was not something expected from
previous data in p+p collisions, although there are data on $\left< p_{T} \right>$ fluctuations 
from STAR~\cite{star_pTfluc} and PHENIX~\cite{phenix_pTfluc} 
that have been interpreted as evidence for similarly large clusters in Au+Au collisions~\cite{cluster_polish}.
The production of jets is a natural mechanism to induce clustering phenomena, 
although one would expect jets to lead to a smaller $\delta$ than isotropic decay. It is also possible that 
additional correlation sources, such as dynamical fluctuation of the $dN/d\eta$ distribution
event-by-event, may modify the two-particle correlations in a way that leads to an increase of the 
observed effective cluster size. The fact that $\delta^{\rm |\eta|<\infty}$
in peripheral A+A collisions is larger than p+p, and far exceeds the value expected for isotropic decay, 
also begs the question as to how the cluster decays are ``elongated'' in phase space. Finally, it is observed
that cluster parameters in central events approach values measured in p+p collisions, while those 
in peripheral events are substantially higher --- almost a factor of two in terms of $K_{\rm eff}-1$ (which is equivalent
to the so-called ``conditional yield'' in analyses involving high $p_{T}$ triggered hadrons).
Overall, more theoretical insights are needed to understand these surprising 
features of two-particle correlations in heavy ion collisions.

\section{Summary and Conclusion}
\label{conclusion}

The two-particle correlation function for inclusive charged particles has been extensively studied 
over a broad range in $\Delta\eta$ and $\Delta\phi$ in p+p, Cu+Cu and Au+Au collisions at 
$\sqrt{s_{_{NN}}}$ = 200~GeV. In particular, it has been shown that the correlation functions 
in heavy ion collisions are not very different from those found in p+p, allowing a 
similar interpretation in terms of clusters. In this approach, 
multiple particles are understood to be emitted close together in phase space, with a typical effective cluster 
size of 2.5-3.5 charged particles in p+p collisions. The correlation functions 
in A+A show a non-trivial decrease in effective cluster 
size with increasing centrality, and a surprising geometric scaling between Cu+Cu and Au+Au collisions. 
Analysis of near- and away-side clusters provides additional information on the details of the 
cluster properties. Extrapolating the measured cluster parameters to the full 
phase space using an independent cluster model as well as other dynamical models
such as PYTHIA, HIJING and AMPT, the effective cluster size and width increase in magnitude to a level which seems 
to challenge most conventional scenarios of the hadronization process. Clearly, more experimental
and theoretical work will be needed to understand these novel aspects of heavy ion collisions. 

%
%
%
%
This work was partially supported by U.S. DOE grants 
DE-AC02-98CH10886,
DE-FG02-93ER40802, 
DE-FG02-94ER40818,  
DE-FG02-94ER40865, 
DE-FG02-99ER41099, and
DE-AC02-06CH11357, by U.S. 
NSF grants 9603486, 
0072204,            
and 0245011,        
by Polish MNiSW grant N N202 282234 (2008-2010),
by NSC of Taiwan Contract NSC 89-2112-M-008-024, and
by Hungarian OTKA grant (F 049823).


\begin{thebibliography}{99}
\bibitem{UA5_3energy} R.\ E.\ Ansorge {\it et al.}, UA5 Collaboration, Z.\ Phys. - Particle and Fields \ {\bf C 37}, 191 (1988).  
\bibitem{cluster_model} F. Henyey, Phys.\ Lett.\ {\bf B 45}, 469 (1973),\\ 
                        E.\ L.\ Berger, Nucl.\ Phys.\ {\bf B 85}, 61 (1975),\\       
                        J.\ L.\ Meunier, G.\ Plaut, Nucl.\ Phys.\ {\bf B 87}, 74 (1975),\\ 
                        C.\ Michael, Nucl.\ Phys.\ {\bf B 103}, 296 (1976). 
\bibitem{ISR_twolowenergy} K.\ Eggert {\it et al.}, Nucl.\ Phys.\ {\bf B 86}, 201 (1975).  
\bibitem{ISR_63GeV} D.\ Drijard {\it et al.}, Nucl.\ Phys.\ {\bf B 155}, 269 (1979). 
\bibitem{cluster_fit} A.\ Morel and G.\ Plaut, Nucl.\ Phys.\ {\bf B 78}, 541 (1974).  
\bibitem{phobos_pp} B.\ Alver {\it et al.}, Phys.\ Rev.\ {\bf C 75}, 054913 (2007). 
\bibitem{AAcluster_prediction} L.\ J.\ Shi and S. Jeon, Phys.\ Rev.\ {\bf C 72}, 034904 (2003). 
\bibitem{phobos_detector}  B.\ B.\ Back {\it et al.}, Nucl.\ Inst.\ Meth.\ {\bf A 499}, 603 (2003). 
\bibitem{hits_paper} B.\ B.\ Back {\it et al.}, Phys.\ Rev.\ Lett.\ {\bf 87}, 102303 (2001),\\
                     Pradeep Sarin, PhD thesis, Massachusetts Institute of Technology (2003).\\
                     Wei Li, PhD thesis, Massachusetts Institute of Technology (2009).
\bibitem{CF_flow} N.\ Borghini, P.\ M.\ Dinh, and J.\-Y.\ Ollitrault, Phys.\ Rev.\ {\bf C 63}, 054906 (2001).
\bibitem{AMPT} Z.\ W.\ Lin, C.\ M.\ Ko, B.\ A.\ Li, B.\ Zhang, and S.\ Pal, Phys.\ Rev.\ {\bf C 72}, 064901 (2005).
\bibitem{phobos_dNdeta_AuAu}  B.\ B.\ Back {\it et al.}, Phys.\ Rev.\ Lett.\ {\bf 91}, 052303 (2003). 
\bibitem{phobos_dNdeta_CuCu}  B.\ Alver {\it et al.}, arXiv:0709.4008v1 (2007). 
\bibitem{star_meanpT}  S.\ Adler {\it et al.}, Phys.\ Rev.\ {\bf C 69}, 034909 (2004). 
\bibitem{PYTHIA} T.\ Sjostrand, S.\ Mrenna, P.\ Skands, JHEP.\ {\bf 0605}, 026 (2006). Version 6.325, single diffractive process excluded. 
\bibitem{HIJING} M.\ Gyulassy and X.\ N.\ Wang, Comput.\ Phys.\ Commun.\ {\bf B 83}, 307 (1994). Version 1.383, single diffractive process excluded. 
\bibitem{Therminator} A.\ Kisiel, T.\ Taluc, W.\ Broniowski and W.\ Florkowski, 
Comput.\ Phys.\ Commun.\ {\bf 174}, 669 (2006) [arXiv:nucl-th/0504047]. Calculation by A. Kisiel, private communication.
\bibitem{star_pTfluc} J.\ Adams {\it et al.}, Phys.\ Rev.\ {\bf C 71}, 064906 (2005),\\
                      J.\ Adams {\it et al.}, Phys.\ Rev.\ {\bf C 72}, 044902 (2005).
\bibitem{phenix_pTfluc} K.\ Adcox {\it et al.}, Phys. Rev.\ {\bf C 66}, 024901 (2002), \\
                        S.\ S.\ Adler {\it et al.}, Phys.\ Rev.\ Lett.\ {\bf 93} 092301 (2004).  
\bibitem{cluster_polish} W.\ Broniowski, P.\ Bozek, W.\ Florkowski and B.\ Hiller, PoS {\bf CFRNC2006}, 020 (2006) [arXiv:nucl-th/0611069].                   

\end{thebibliography}
\end{document}